\documentclass[12pt,a4paper]{article}

\usepackage[T1]{fontenc}
\usepackage[utf8]{inputenc}
\usepackage[
  top=1.25in,
  bottom=1.25in,
  left=0.90in,
  right=0.90in
]{geometry}
\usepackage{amsmath,amssymb,amsthm}
\usepackage{appendix}
\usepackage{enumitem}
\usepackage{etoolbox}
\usepackage{microtype}
\microtypesetup{nopatch=footnote}
\usepackage{placeins}
\usepackage{tikz,pgfplots}
\pgfplotsset{compat=1.18}
\usepackage{csquotes}
\usepackage[
  backend=biber,
  style=authoryear,
  sortcites=true,
  maxcitenames=3,
  maxbibnames=99,
  doi=false,
  url=false,
  isbn=false
]{biblatex}
\DeclareDelimFormat{nameyeardelim}{\addcomma\space}
\usepackage[hidelinks,hyperfootnotes=false]{hyperref}
\hypersetup{
  pdftitle={When Public Information Helps Consumers},
  pdfauthor={Keita Kuwahara},
  pdfsubject={Public information, monopoly pricing, and consumer surplus},
  pdfkeywords={
    public information,
    monopoly pricing,
    consumer surplus,
    information disclosure,
    virtual values
  }
}
\usepackage{setspace}
\AtBeginEnvironment{appendices}{

  \setcounter{lemma}{0}
}

\newtheorem{lemma}{Lemma}
\newtheorem{proposition}{Proposition}
\newtheorem{theorem}{Theorem}
\newtheorem{corollary}{Corollary}
\theoremstyle{definition}
\newtheorem{example}{Example}
\theoremstyle{plain}

\newcommand{\E}{\mathbb{E}}
\newcommand{\dd}{\,\mathrm{d}}
\newcommand{\CS}{\mathrm{CS}}
\newcommand{\PS}{\mathrm{PS}}
\newcommand{\TS}{\mathrm{TS}}

\title{When Public Information Helps Consumers\thanks{
I thank Yu Awaya, Ming Hu, Shota Ichihashi, Michihiro Kandori, Fuhito Kojima, Joosung Lee, Yichuan Lou, Stephen Morris, Shunya Noda, Daisuke Oyama, Satoru Takahashi, Yuichi Yamamoto, Takuro Yamashita, and participants at the 2026 Game Theory Workshop, the 2026 Spring Meeting of the Japanese Economic Association, and the 18th Meeting of the Society for Social Choice and Welfare for helpful comments. This paper was previously circulated under the title ``When Sellers Are Uncertain about Quality.'' All remaining errors are my own.
}}
\author{Keita Kuwahara\thanks{Graduate School of Economics, The University of Tokyo.
Email: \texttt{kuke0303@g.ecc.u-tokyo.ac.jp}}}
\date{\today}

\begin{document}

\maketitle

\doublespacing

\begin{abstract}
Does more information benefit consumers when a monopolist can adjust its price? Although information helps consumers evaluate a product, \textcite{schlee1996value} showed that public disclosure can reduce consumer surplus. We characterize exactly when the opposite occurs. More informative disclosure increases consumer surplus if and only if a weighted combination of the slope and curvature of Myerson's virtual type is nonnegative. Producer surplus always increases, while a weaker condition characterizes when total surplus increases. In particular, information benefits consumers whenever the virtual type is convex. Even when convexity fails, the same conclusion holds if monopoly output is sufficiently small.
\end{abstract}


\newpage

\section{Introduction}

In many markets, public information and posted prices now change almost simultaneously. A weather service announces that bad weather is likely. The trip becomes less attractive to travelers, demand for hotel rooms falls, and hotels may respond by lowering their rates. A food-delivery platform announces a shorter expected delivery time. The meal becomes more attractive to consumers, demand rises, and the seller may respond by charging a higher price. These examples illustrate that information helps consumers evaluate a purchase, but it also enables firms to adjust their prices.

This issue is especially important in increasingly data-rich markets, where information is released quickly, firms can track demand closely, and prices can adjust almost immediately. Disclosure therefore affects not only what consumers know, but also the firm's pricing and sales decisions. Evaluating disclosure while holding prices fixed may thus miss its central welfare effect.

\textcite{schlee1996value} gave the central warning: even when consumers and a monopolist receive the same information, consumers can lose because the monopolist changes its price. The paper also obtained a positive result for linear demand and linear cost. This raises a natural question: how can we tell whether information helps consumers in a general market, and is the positive result limited to linear demand?

This paper provides an exact demand-side test.
After a public message is observed by both sides, the monopolist sets a price. Consumers have private types that determine their willingness to pay, and each consumer demands at most one unit.
We characterize the necessary and sufficient conditions under which more informative disclosure increases consumer surplus or total surplus. These conditions are expressed in terms of the virtual type introduced by \textcite{myerson1981optimal}: they require a combination of its slope and curvature to be nonnegative.
Producer surplus, by contrast, always increases without any additional condition.

The positive result applies broadly. If the virtual type is convex, more information increases both consumer surplus and total surplus. This condition holds for the uniform distribution, which corresponds to linear demand, as well as for every Beta$(\alpha,\beta)$ distribution with $\alpha\leq 1$, so the result is not limited to linear demand or isolated functional-form examples. Even when the virtual type is not convex, more information still increases both surplus measures whenever monopoly output is sufficiently small.

The intuition for why more information benefits consumers in a broad class of cases is as follows. Favorable news raises willingness to pay and, even after the monopolist adjusts its price, expands output. The resulting additional trades therefore occur when consumers' values are high. Unfavorable news, by contrast, reduces output when values are low, so the resulting loss in consumer surplus tends to be smaller. Our main condition characterizes exactly when the former positive effect dominates the latter negative effect.

\paragraph{Related literature.}
We build on \citeauthor{schlee1996value}'s (\citeyear{schlee1996value}) observation that public information can hurt consumers.
The paper establishes this possibility, derives cost-side conditions under which consumers may lose, and provides a positive benchmark with linear demand and linear cost. Holding marginal cost constant, we instead derive primitive necessary and sufficient demand-side conditions under which consumer surplus and total surplus are nondecreasing under every Blackwell improvement. We also identify broad nonlinear classes of demand for which these conditions hold.

A particularly close paper in terms of game timing and market structure is \textcite{agrawal2026simple}, who study public quality disclosure followed by signal-contingent monopoly pricing with heterogeneous buyer types. Their objective is to obtain robust seller-revenue guarantees for simple quantile-partition policies across market environments, whereas we fix the market environment and characterize consumer- and total-surplus monotonicity under every Blackwell improvement.
They also establish convexity of the seller's optimal posted-price revenue and extend the revenue optimality of full disclosure to an additional constant-marginal-rate-of-substitution class.
Accordingly, producer-surplus monotonicity serves mainly as a benchmark in our analysis; our principal contribution is the primitive characterization of consumer- and total-surplus monotonicity.

Related principal-payoff results include \textcite{ottaviani2001value}, who study affiliated public information for a nonlinear-pricing monopolist selling to a single buyer, and \textcite{yamashita2018optimal}, who studies public disclosure chosen before optimal mechanism design. \textcite{sugaya2026collusion} study selective disclosure by an information intermediary in repeated Bertrand competition and show that improving the intermediary's information accuracy reduces expected consumer surplus whenever the corresponding monopoly comparison does.

A related literature studies consumer data and price discrimination. \textcite{bergemann2015limits} characterize the full set of consumer--producer surplus pairs attainable when a monopolist segments a market using information about consumer tastes. \textcite{farboodi2025good} characterize when refinements of consumer-data-based market segmentation monotonically raise or lower a weighted sum of consumer and producer surplus. \textcite{ali2023voluntary} study consumer-controlled disclosure under personalized pricing, while \textcite{ichihashi2020online} studies disclosure to a multiproduct seller that uses information for product recommendations and pricing. These papers concern information about individual consumers or market composition and allow prices to vary across consumers or market segments. By contrast, our signal concerns a common market state, is observed publicly, and is followed by one posted price common to all consumers after each realized message.

Other papers study the design of information received by consumers. In a competitive environment, \textcite{armstrong2022consumer} characterize firm- and consumer-optimal private signals about consumers' product preferences. \textcite{bonatti2026theory} develop a general curvature-based framework for the individual welfare effects of privacy, viewing a loss of privacy as more precise observers' beliefs about an individual's type. Our curvature argument is related in spirit, but it concerns consumer surplus under public information about a common market state rather than privacy about an individual's characteristics.

Other work studies disclosure jointly with endogenous product or contract design. \textcite{ma2014public} study public reports of multidimensional quality when a monopolist chooses a price--quality menu. \textcite{salamat2025trial} studies a seller that screens privately informed consumers with a menu of trial experiments, trial prices, and product prices, and provides conditions for full surplus extraction. In that model, consumers select personalized trial plans that reveal idiosyncratic match quality; in ours, all consumers observe the same public signal about a common state, and the monopolist responds with one posted price common to all consumers. Methodologically, our formulation in terms of Bayes-plausible posterior beliefs is related to the information-design literature \parencite{rayo2010optimal,kamenica2011bayesian, bergemann2016information}.

Section~\ref{sec:model} presents the model, Section~\ref{sec:results} gives the results, and Section~\ref{sec:examples} gives two examples. All proofs are in the appendix.

\section{Model}\label{sec:model}

\subsection{Consumers and the Market State}

There is a continuum of consumers with total mass one. A monopolist sells one good, and each consumer buys at most one unit. A consumer's private type is $\theta\in[0,\bar\theta]$. Types follow a distribution function $F$ with a positive, twice continuously differentiable density $f$ on $(0,\bar\theta)$. Define
\[
  r(\theta)=\frac{1-F(\theta)}{f(\theta)},
  \qquad
  \varphi(\theta)=\theta-r(\theta).
\]
The function $r$ is the \emph{inverse hazard rate}. Following \textcite{myerson1981optimal}, $\varphi$ is the \emph{virtual type}. The distribution is \emph{regular} if
\[
  \varphi'(\theta)>0
  \qquad
  \text{whenever }\varphi(\theta)\ge0.
\]
We assume regularity.
By Lemma~\ref{lem:tail}, $\varphi$ has a unique zero
$\theta_0\in(0,\bar\theta)$, is strictly increasing on
$[\theta_0,\bar\theta)$, and converges to $\bar\theta$ as
$\theta\uparrow\bar\theta$. We denote by
\[
  \varphi^{-1}\colon [0,\bar\theta)\to[\theta_0,\bar\theta)
\]
the inverse of $\varphi$ restricted to $[\theta_0,\bar\theta)$.

The market state \(\omega\) lies in a compact metric space \(\Omega\).
Let $\Delta(\Omega)$ denote the set of Borel probability distributions on $\Omega$.
The common prior $\mu_0\in\Delta(\Omega)$ has full support. A type-$\theta$ consumer's value is \(v(\theta,\omega)\).
The function \(v\) is continuous, and \(v(\cdot,\omega)\) is affine and strictly increasing for every \(\omega\).
Thus, there are continuous functions \(a:\Omega\to\mathbb{R}_{>0}\) and \(b:\Omega\to\mathbb{R}\) such that
\[
  v(\theta,\omega)=a(\omega)\theta+b(\omega).
\]
If the consumer buys at price $p$, utility is $v(\theta,\omega)-p$; if the consumer does not buy, utility is zero. Types and the state are independent.

The monopolist has constant marginal cost $c\ge0$. We assume that
\[
  \max_{\omega\in \Omega}v(0,\omega)=\max_{\omega\in \Omega}b(\omega)\le c.
\]
Thus, serving the lowest type is not profitable in any state.

\subsection{Information}

An \emph{information policy} is a pair $(S,\pi)$, where $S$ is a set of public messages and \(\pi(\mathrm ds\mid\omega)\) is a Markov kernel from \(\Omega\) to \(S\). Upon observing a message $s\in S$, the monopolist and consumers update their common prior according to Bayes' rule and share a posterior belief $\mu_s\in\Delta(\Omega)$.
Define
\[
  a(\mu)=\int_\Omega a(\omega)\,\mu(\dd\omega),
  \qquad
  b(\mu)=\int_\Omega b(\omega)\,\mu(\dd\omega).
\]
Write $(a_s,b_s)=(a(\mu_s),b(\mu_s))$ and call it the \emph{posterior demand pair}. Conditional on message \(s\), a type-\(\theta\) consumer's expected value is
\[
a_s\theta+b_s.
\]

Let
\[
  K
  =\{(a(\mu),b(\mu)):\mu\in\Delta(\Omega)\}.
\]
Every posterior pair lies in $K$.
By construction, \(K\) is the convex hull of the image of the continuous map
\(\omega\mapsto(a(\omega),b(\omega))\). Hence, \(K\) is compact and convex.

An information policy is \emph{more informative} than another if the latter can be obtained by garbling the former \parencite{blackwell1953equivalent}. \emph{Full disclosure} reveals the state, while \emph{no disclosure} always sends the same message.

The timing is:
\begin{enumerate}[label=(\arabic*)]
  \item An information policy $(S,\pi)$ is fixed.
  \item Nature draws \(\omega\sim\mu_0\), and then a public message \(s\) is drawn according to $\pi(\cdot\mid \omega)$.
  \item The monopolist observes $s$ and posts a price.\footnote{Restricting attention to posted prices is without loss of optimality among incentive-compatible and individually rational direct mechanisms. Conditional on a message, types are one-dimensional and expected values are increasing in type. Regularity implies that the optimal allocation serves all consumers above a cutoff and can be implemented by a posted price; see \textcite{myerson1981optimal}.}
  \item Each consumer observes the same message and the price and buys if her expected value is at least the price.
\end{enumerate}

\subsection{Pricing and Surplus}

Fix a posterior pair $(a,b)\in K$. If the monopolist serves all types above a \emph{cutoff} $x\in[0,\bar\theta]$, the implementing price is $p=ax+b$. The convention $x=\bar\theta$ means no trade. Define the \emph{break-even type}
\[
  t(a,b)=\frac{c-b}{a}\ge0.
\]
The monopolist solves
\begin{equation}
\label{eq:ps}
\max_{x\in[0,\bar\theta]}
(ax+b-c)[1-F(x)].
\end{equation}

\begin{lemma}[Monopoly cutoff]\label{lem:cutoff}
For every $(a,b)\in K$, the unique optimal cutoff is continuous in $(a,b)$ and equals
\begin{equation}
\label{eq:cutoff}
  x(a,b)=
  \begin{cases}
    \varphi^{-1}\!\left(t(a,b)\right), & t(a,b)<\bar\theta,\\[1mm]
    \bar\theta, & t(a,b)\ge\bar\theta.
  \end{cases}
\end{equation}
\end{lemma}
Under regularity, the cutoff $x(a,b)$ decreases as either $a$ or $b$ increases. Hence, \emph{output},
\[
  q(a,b)=1-F\bigl(x(a,b)\bigr),
\]
increases with both $a$ and $b$. Thus, good news expands output, whereas bad news contracts it.

Let
\[
  x_L=\min_{(a,b)\in K}x(a,b),
  \qquad
  x_H=\max_{(a,b)\in K}x(a,b).
\]
Write $x=x(a,b)$ and $t=t(a,b)$. Producer surplus, consumer surplus, and total surplus are
\begin{equation}
\label{eq:surplus}
\begin{aligned}
  \PS(a,b)&=a(x-t)[1-F(x)],\\
  \CS(a,b)&=a\int_x^{\bar\theta}(\theta-x)f(\theta)\dd\theta,\\
  \TS(a,b)&=a\int_x^{\bar\theta}(\theta-t)f(\theta)\dd\theta.
\end{aligned}
\end{equation}
The expected producer, consumer, and total surpluses under an information policy are
\[
\E[\PS(a_s,b_s)],\qquad
\E[\CS(a_s,b_s)],\qquad
\E[\TS(a_s,b_s)],
\]
respectively.

\section{Results}\label{sec:results}

A surplus measure is \emph{increasing in information} if its expected value weakly increases whenever an information policy is replaced by a more informative one.

\begin{proposition}[Producer surplus]\label{prop:producer}
Producer surplus is increasing in information.
\end{proposition}

Under the more informative policy, the monopolist can always ignore the additional information. In particular, it can privately simulate the garbling that generates the less informative message and use the cutoff that would be optimal under that message. This strategy yields the same expected producer surplus as the less informative policy. The monopolist can therefore do weakly better by using the additional information and reoptimizing.

The more subtle question is how the resulting price adjustment affects consumer surplus.

\begin{theorem}[Consumer surplus]\label{thm:consumer}
Consumer surplus is increasing in information if and only if
\begin{equation}
\label{eq:cs-condition}
  \varphi'(\theta)+r(\theta)\varphi''(\theta)\ge0
  \qquad
  \text{for every }\theta\in(x_L,x_H).
\end{equation}
\end{theorem}

The interval $(x_L,x_H)$ is the range over which the optimal cutoff can vary. Condition~\eqref{eq:cs-condition} can be interpreted as requiring a weighted combination of the slope and curvature of the virtual type $\varphi$ to remain nonegative throughout the relevant range, so that the monopolist's price response to information is not too unfavorable to consumers.

To see the role of this condition, consider a coupling of the more and less informative policies. Let $s_1$ and $s_0$ denote the messages generated by the respective policies, and let
\[
  k_i=(a_{s_i},b_{s_i}),
  \qquad i\in\{0,1\},
\]
be the corresponding posterior demand pairs. Since the less informative policy is a garbling of the more informative one,
\[
  k_0=\E[k_1\mid s_0].
\]
Greater informativeness therefore generates a conditional mean-preserving spread in posterior demand. It follows that consumer surplus is increasing in information whenever $\CS(a,b)$ is convex in $(a,b)$. In the appendix, we show that this convexity is equivalent to condition~\eqref{eq:cs-condition}.

Similarly, we next characterize when total surplus is increasing in information.

\begin{theorem}[Total surplus]\label{thm:total}
Total surplus is increasing in information if and only if
\begin{equation}
\label{eq:ts-condition}
  \varphi'(\theta)+\bigl(\varphi'(\theta)\bigr)^2
  +r(\theta)\varphi''(\theta)\ge0
  \qquad
  \text{for every }\theta\in(x_L,x_H).
\end{equation}
\end{theorem}

Condition~\eqref{eq:ts-condition} is weaker than condition~\eqref{eq:cs-condition}, as it includes the additional nonnegative term \(\bigl(\varphi'(\theta)\bigr)^2\). By Proposition~\ref{prop:producer}, producer surplus is increasing in information. Therefore, even if consumer surplus decreases, the resulting gain in producer surplus may be large enough to increase total surplus.

Under regularity, Theorems~\ref{thm:consumer} and~\ref{thm:total} immediately yield the following useful sufficient condition.

\begin{corollary}[Convex virtual type]\label{cor:convex-virtual}
If $\varphi$ is convex on $(x_L,x_H)$, then producer surplus, consumer surplus, and total surplus are increasing in information.
\end{corollary}

Convexity of the virtual type holds for a broad and economically useful family of distributions. As shown in Example~\ref{ex:beta-family}, it is satisfied by many beta distributions, including the uniform distribution.
Even when this condition fails, the following proposition shows that information benefits consumers whenever optimal output is sufficiently small.

\begin{proposition}[Small-output guarantee]\label{prop:small-output}
Suppose $\varphi''$ is bounded below on a left-neighborhood of $\bar\theta$. Then there exists \(\bar q>0\), depending only on \(F\), such that if
\[
  q(a,b)\le\bar q
  \qquad
  \text{for every }(a,b)\in K,
\]
consumer surplus and total surplus are increasing in information.
\end{proposition}

The condition that $\varphi''$ is bounded below on a left-neighborhood of $\bar\theta$ is satisfied by many standard distributions, including uniform, beta, and truncated normal distributions. Proposition~\ref{prop:small-output} therefore has a simple implication: if the monopolist's optimal output is sufficiently small under every belief, then more informative disclosure increases producer surplus, consumer surplus, and total surplus.

To see the intuition, small optimal output means that every optimal cutoff lies close to the highest type $\bar\theta$. By Lemma~\ref{lem:tail},
\[
    \lim_{\theta\uparrow\bar\theta} r(\theta)=0.
\]
Thus, near $\bar\theta$, the potentially adverse term involving $\varphi''$ in condition~\eqref{eq:cs-condition} becomes negligible. Regularity then ensures that the condition is satisfied, so more information benefits consumers.

\section{Examples}\label{sec:examples}

\begin{example}[A broad Beta family]\label{ex:beta-family}
Let $\bar\theta=1$, and suppose that types follow a Beta$(\alpha,\beta)$ distribution with \(\alpha\le1\).
Its density is given by
\[
  f(\theta)=\frac{\theta^{\alpha-1}(1-\theta)^{\beta-1}}
  {\mathrm{B}(\alpha,\beta)}.
\]
Because the virtual type is convex and the distribution is regular, Corollary~\ref{cor:convex-virtual} applies. Thus, more information increases producer surplus, consumer surplus, and total surplus for any state space \(\Omega\), value function \(v\), and marginal cost \(c\).
This family includes the uniform distribution as the special case \(\alpha=\beta=1\), which corresponds to the linear-demand setting studied by \textcite{schlee1996value}. It also encompasses many nonlinear demand curves.
\end{example}

\begin{example}[Beta$(3,3)$]\label{ex:beta33}
Let consumer types follow a $\operatorname{Beta}(3,3)$ distribution on $[0,1]$, let $c=1$, and suppose that
\[
v(\theta,\omega)=\theta+\omega.
\]
There are two equally likely states, $\Omega=\{\omega_L,\omega_H\}$, with $\omega_L<\omega_H$.

First set
\[
(\omega_L,\omega_H)=(0.6,0.8).
\]
In this case, the small-output bound in Proposition~\ref{prop:small-output} can be taken to be approximately $0.612267$, whereas the maximum optimal output is approximately $0.530669$. Proposition~\ref{prop:small-output} therefore applies. Consequently, full disclosure increases and maximizes producer surplus, consumer surplus, and total surplus.

Next set
\[
(\omega_L,\omega_H)=(0.8,1).
\]
The maximum optimal output is approximately $0.685794$, so the small-output guarantee no longer applies. In fact, full disclosure reduces consumer surplus relative to no disclosure. A consumer-optimal policy uses two messages, $\mathsf{G}$ and $\mathsf{B}$. In state $\omega_H$, it sends $\mathsf{G}$ with probability one. In state $\omega_L$, it sends $\mathsf{B}$ with probability $0.644465$ and $\mathsf{G}$ otherwise.

Figure~\ref{fig:beta33} shows the full set of attainable surplus outcomes in the two cases.
\end{example}

\tikzset{
  surplus label/.style={
    font=\scriptsize,
    fill=white,
    fill opacity=0.9,
    text opacity=1,
    inner sep=1.5pt
  }
}
\begin{figure}[!htbp]
  \centering
  \begin{minipage}[t]{0.5\textwidth}
    \centering
    \begin{tikzpicture}
      \begin{axis}[
        width=7.1cm,
        height=5.2cm,
        xmin=-0.08,
        xmax=1.72,
        ymin=-0.15,
        ymax=5.30,
        axis lines=left,
        axis on top,
        xlabel={$10^3\Delta\CS$},
        ylabel={$10^3\Delta\PS$},
        xtick={0,0.5,1,1.5},
        ytick={0,1,2,3,4,5},
        tick label style={font=\small},
        label style={font=\small},
        clip=false
      ]
        \addplot[draw=none,fill=black!10] coordinates {
          (0.00000000,0.00000000) (0.00900274,0.02535545) (0.01808389,0.05096289) (0.02724446,0.07682606)
          (0.03648543,0.10294879) (0.04580785,0.12933498) (0.05521274,0.15598861) (0.06470116,0.18291374)
          (0.07427419,0.21011452) (0.08393291,0.23759518) (0.09367843,0.26536001) (0.10351189,0.29341344)
          (0.11343441,0.32175994) (0.12344718,0.35040411) (0.13355136,0.37935063) (0.14374817,0.40860427)
          (0.15403881,0.43816991) (0.16442454,0.46805254) (0.17490661,0.49825724) (0.18548631,0.52878922)
          (0.19616493,0.55965378) (0.20694382,0.59085635) (0.21782430,0.62240246) (0.22880776,0.65429777)
          (0.23989558,0.68654809) (0.25108918,0.71915931) (0.26239000,0.75213748) (0.27379951,0.78548879)
          (0.28531918,0.81921955) (0.29695053,0.85333622) (0.30869510,0.88784540) (0.32055445,0.92275386)
          (0.33253016,0.95806851) (0.34462387,0.99379641) (0.35683720,1.02994480) (0.36917184,1.06652109)
          (0.38162946,1.10353285) (0.39421181,1.14098783) (0.40692064,1.17889398) (0.41975773,1.21725941)
          (0.43272490,1.25609246) (0.44582399,1.29540164) (0.45905687,1.33519568) (0.47242545,1.37548352)
          (0.48593167,1.41627432) (0.49957750,1.45757748) (0.51336493,1.49940259) (0.52729602,1.54175954)
          (0.54137281,1.58465841) (0.55559742,1.62810958) (0.56997197,1.67212366) (0.58449865,1.71671156)
          (0.59917965,1.76188443) (0.61401721,1.80765376) (0.62901362,1.85403130) (0.64417118,1.90102913)
          (0.65949224,1.94865962) (0.67497918,1.99693550) (0.69063444,2.04586982) (0.70646045,2.09547599)
          (0.72245973,2.14576778) (0.73863480,2.19675933) (0.75498824,2.24846516) (0.77152265,2.30090022)
          (0.78824067,2.35407983) (0.80514500,2.40801978) (0.82223834,2.46273627) (0.83952347,2.51824599)
          (0.85700316,2.57456606) (0.87468025,2.63171414) (0.89255760,2.68970837) (0.91063813,2.74856742)
          (0.92892476,2.80831052) (0.94742046,2.86895743) (0.96612823,2.93052855) (0.98505111,2.99304483)
          (1.00419217,3.05652790) (1.02355450,3.12100001) (1.04314121,3.18648411) (1.06295546,3.25300383)
          (1.08300041,3.32058355) (1.10327925,3.38924841) (1.12379520,3.45902433) (1.14455147,3.52993804)
          (1.16555131,3.60201713) (1.18679796,3.67529008) (1.20829466,3.74978628) (1.23004466,3.82553607)
          (1.25205122,3.90257080) (1.27431755,3.98092283) (1.29684687,4.06062562) (1.31964239,4.14171374)
          (1.34270725,4.22422293) (1.36604458,4.30819013) (1.38965746,4.39365357) (1.41354890,4.48065278)
          (1.43772185,4.56922868) (1.46217918,4.65942361) (1.48692366,4.75128143) (1.51195797,4.84484752)
          (1.53728464,4.94016894) (1.50216683,4.84081760) (1.46790785,4.74345233) (1.43447866,4.64801555)
          (1.40185147,4.55445181) (1.36999965,4.46270772) (1.33889765,4.37273186) (1.30852101,4.28447466)
          (1.27884625,4.19788836) (1.24985085,4.11292692) (1.22151320,4.02954593) (1.19381253,3.94770255)
          (1.16672893,3.86735543) (1.14024324,3.78846467) (1.11433708,3.71099177) (1.08899274,3.63489950)
          (1.06419324,3.56015195) (1.03992221,3.48671437) (1.01616391,3.41455322) (0.99290321,3.34363605)
          (0.97012550,3.27393149) (0.94781677,3.20540918) (0.92596347,3.13803978) (0.90455259,3.07179487)
          (0.88357154,3.00664695) (0.86300824,2.94256940) (0.84285099,2.87953644) (0.82308853,2.81752309)
          (0.80371000,2.75650516) (0.78470491,2.69645919) (0.76606313,2.63736247) (0.74777488,2.57919295)
          (0.72983072,2.52192927) (0.71222153,2.46555070) (0.69493849,2.41003714) (0.67797309,2.35536908)
          (0.66131707,2.30152758) (0.64496248,2.24849426) (0.62890161,2.19625128) (0.61312701,2.14478130)
          (0.59763147,2.09406750) (0.58240799,2.04409351) (0.56744982,1.99484343) (0.55275042,1.94630184)
          (0.53830345,1.89845371) (0.52410276,1.85128444) (0.51014240,1.80477984) (0.49641661,1.75892611)
          (0.48291980,1.71370981) (0.46964653,1.66911787) (0.45659157,1.62513757) (0.44374981,1.58175655)
          (0.43111629,1.53896275) (0.41868623,1.49674444) (0.40645496,1.45509020) (0.39441797,1.41398889)
          (0.38257086,1.37342969) (0.37090936,1.33340203) (0.35942934,1.29389562) (0.34812678,1.25490044)
          (0.33699776,1.21640671) (0.32603849,1.17840490) (0.31524528,1.14088572) (0.30461453,1.10384011)
          (0.29414277,1.06725922) (0.28382659,1.03113445) (0.27366269,0.99545738) (0.26364787,0.96021980)
          (0.25377901,0.92541371) (0.24405305,0.89103128) (0.23446705,0.85706489) (0.22501812,0.82350708)
          (0.21570346,0.79035059) (0.20652035,0.75758830) (0.19746611,0.72521328) (0.18853817,0.69321875)
          (0.17973400,0.66159810) (0.17105115,0.63034483) (0.16248721,0.59945265) (0.15403985,0.56891536)
          (0.14570681,0.53872693) (0.13748585,0.50888145) (0.12937482,0.47937314) (0.12137161,0.45019637)
          (0.11347417,0.42134561) (0.10568047,0.39281546) (0.09798858,0.36460064) (0.09039658,0.33669599)
          (0.08290260,0.30909644) (0.07550483,0.28179705) (0.06820149,0.25479298) (0.06099085,0.22807949)
          (0.05387122,0.20165194) (0.04684094,0.17550580) (0.03989840,0.14963661) (0.03304203,0.12404003)
          (0.02627028,0.09871178) (0.01958166,0.07364771) (0.01297469,0.04884371) (0.00644794,0.02429578)
          (0.00000000,0.00000000)
        } \closedcycle;
        \addplot[black,thick] coordinates {
          (0.00000000,0.00000000) (0.00900274,0.02535545) (0.01808389,0.05096289) (0.02724446,0.07682606)
          (0.03648543,0.10294879) (0.04580785,0.12933498) (0.05521274,0.15598861) (0.06470116,0.18291374)
          (0.07427419,0.21011452) (0.08393291,0.23759518) (0.09367843,0.26536001) (0.10351189,0.29341344)
          (0.11343441,0.32175994) (0.12344718,0.35040411) (0.13355136,0.37935063) (0.14374817,0.40860427)
          (0.15403881,0.43816991) (0.16442454,0.46805254) (0.17490661,0.49825724) (0.18548631,0.52878922)
          (0.19616493,0.55965378) (0.20694382,0.59085635) (0.21782430,0.62240246) (0.22880776,0.65429777)
          (0.23989558,0.68654809) (0.25108918,0.71915931) (0.26239000,0.75213748) (0.27379951,0.78548879)
          (0.28531918,0.81921955) (0.29695053,0.85333622) (0.30869510,0.88784540) (0.32055445,0.92275386)
          (0.33253016,0.95806851) (0.34462387,0.99379641) (0.35683720,1.02994480) (0.36917184,1.06652109)
          (0.38162946,1.10353285) (0.39421181,1.14098783) (0.40692064,1.17889398) (0.41975773,1.21725941)
          (0.43272490,1.25609246) (0.44582399,1.29540164) (0.45905687,1.33519568) (0.47242545,1.37548352)
          (0.48593167,1.41627432) (0.49957750,1.45757748) (0.51336493,1.49940259) (0.52729602,1.54175954)
          (0.54137281,1.58465841) (0.55559742,1.62810958) (0.56997197,1.67212366) (0.58449865,1.71671156)
          (0.59917965,1.76188443) (0.61401721,1.80765376) (0.62901362,1.85403130) (0.64417118,1.90102913)
          (0.65949224,1.94865962) (0.67497918,1.99693550) (0.69063444,2.04586982) (0.70646045,2.09547599)
          (0.72245973,2.14576778) (0.73863480,2.19675933) (0.75498824,2.24846516) (0.77152265,2.30090022)
          (0.78824067,2.35407983) (0.80514500,2.40801978) (0.82223834,2.46273627) (0.83952347,2.51824599)
          (0.85700316,2.57456606) (0.87468025,2.63171414) (0.89255760,2.68970837) (0.91063813,2.74856742)
          (0.92892476,2.80831052) (0.94742046,2.86895743) (0.96612823,2.93052855) (0.98505111,2.99304483)
          (1.00419217,3.05652790) (1.02355450,3.12100001) (1.04314121,3.18648411) (1.06295546,3.25300383)
          (1.08300041,3.32058355) (1.10327925,3.38924841) (1.12379520,3.45902433) (1.14455147,3.52993804)
          (1.16555131,3.60201713) (1.18679796,3.67529008) (1.20829466,3.74978628) (1.23004466,3.82553607)
          (1.25205122,3.90257080) (1.27431755,3.98092283) (1.29684687,4.06062562) (1.31964239,4.14171374)
          (1.34270725,4.22422293) (1.36604458,4.30819013) (1.38965746,4.39365357) (1.41354890,4.48065278)
          (1.43772185,4.56922868) (1.46217918,4.65942361) (1.48692366,4.75128143) (1.51195797,4.84484752)
          (1.53728464,4.94016894)
        };
        \addplot[black,thick] coordinates {
          (0.00000000,0.00000000) (0.00644794,0.02429578) (0.01297469,0.04884371) (0.01958166,0.07364771)
          (0.02627028,0.09871178) (0.03304203,0.12404003) (0.03989840,0.14963661) (0.04684094,0.17550580)
          (0.05387122,0.20165194) (0.06099085,0.22807949) (0.06820149,0.25479298) (0.07550483,0.28179705)
          (0.08290260,0.30909644) (0.09039658,0.33669599) (0.09798858,0.36460064) (0.10568047,0.39281546)
          (0.11347417,0.42134561) (0.12137161,0.45019637) (0.12937482,0.47937314) (0.13748585,0.50888145)
          (0.14570681,0.53872693) (0.15403985,0.56891536) (0.16248721,0.59945265) (0.17105115,0.63034483)
          (0.17973400,0.66159810) (0.18853817,0.69321875) (0.19746611,0.72521328) (0.20652035,0.75758830)
          (0.21570346,0.79035059) (0.22501812,0.82350708) (0.23446705,0.85706489) (0.24405305,0.89103128)
          (0.25377901,0.92541371) (0.26364787,0.96021980) (0.27366269,0.99545738) (0.28382659,1.03113445)
          (0.29414277,1.06725922) (0.30461453,1.10384011) (0.31524528,1.14088572) (0.32603849,1.17840490)
          (0.33699776,1.21640671) (0.34812678,1.25490044) (0.35942934,1.29389562) (0.37090936,1.33340203)
          (0.38257086,1.37342969) (0.39441797,1.41398889) (0.40645496,1.45509020) (0.41868623,1.49674444)
          (0.43111629,1.53896275) (0.44374981,1.58175655) (0.45659157,1.62513757) (0.46964653,1.66911787)
          (0.48291980,1.71370981) (0.49641661,1.75892611) (0.51014240,1.80477984) (0.52410276,1.85128444)
          (0.53830345,1.89845371) (0.55275042,1.94630184) (0.56744982,1.99484343) (0.58240799,2.04409351)
          (0.59763147,2.09406750) (0.61312701,2.14478130) (0.62890161,2.19625128) (0.64496248,2.24849426)
          (0.66131707,2.30152758) (0.67797309,2.35536908) (0.69493849,2.41003714) (0.71222153,2.46555070)
          (0.72983072,2.52192927) (0.74777488,2.57919295) (0.76606313,2.63736247) (0.78470491,2.69645919)
          (0.80371000,2.75650516) (0.82308853,2.81752309) (0.84285099,2.87953644) (0.86300824,2.94256940)
          (0.88357154,3.00664695) (0.90455259,3.07179487) (0.92596347,3.13803978) (0.94781677,3.20540918)
          (0.97012550,3.27393149) (0.99290321,3.34363605) (1.01616391,3.41455322) (1.03992221,3.48671437)
          (1.06419324,3.56015195) (1.08899274,3.63489950) (1.11433708,3.71099177) (1.14024324,3.78846467)
          (1.16672893,3.86735543) (1.19381253,3.94770255) (1.22151320,4.02954593) (1.24985085,4.11292692)
          (1.27884625,4.19788836) (1.30852101,4.28447466) (1.33889765,4.37273186) (1.36999965,4.46270772)
          (1.40185147,4.55445181) (1.43447866,4.64801555) (1.46790785,4.74345233) (1.50216683,4.84081760)
          (1.53728464,4.94016894)
        };
        \addplot[black,only marks,mark=*,mark size=2.1pt]
          coordinates {(0,0) (1.53728464,4.94016894)};
        \node[surplus label,anchor=north west]
          at (axis cs:0.05,0.33) {No disclosure};
        \node[surplus label,anchor=south east]
          at (axis cs:1.48728464,4.94016894) {Full disclosure};
      \end{axis}
    \end{tikzpicture}
    \par\smallskip
    {\small (a) $\Omega=\{0.6,0.8\}$}
  \end{minipage}\hfill
  \begin{minipage}[t]{0.5\textwidth}
    \centering
    \begin{tikzpicture}
      \begin{axis}[
        width=7.1cm,
        height=5.2cm,
        xmin=-6.75,
        xmax=5.85,
        ymin=-0.12,
        ymax=4.22,
        axis lines=left,
        axis on top,
        xlabel={$10^5\Delta\CS$},
        ylabel={$10^3\Delta\PS$},
        xtick={-6,-4,-2,0,2,4},
        ytick={0,1,2,3,4},
        tick label style={font=\small},
        label style={font=\small},
        clip=false
      ]
        \addplot[draw=none,fill=black!10] coordinates {
          (0.00000000,0.00000000) (0.10965277,0.02060405) (0.21917570,0.04140513) (0.32854900,0.06240607)
          (0.43775214,0.08360975) (0.54676382,0.10501910) (0.65556196,0.12663712) (0.76412362,0.14846685)
          (0.87242504,0.17051142) (0.98044153,0.19277398) (1.08814752,0.21525777) (1.19551644,0.23796610)
          (1.30252078,0.26090232) (1.40913195,0.28406986) (1.51532032,0.30747222) (1.62105516,0.33111297)
          (1.72630458,0.35499576) (1.83103554,0.37912430) (1.93521363,0.40350238) (2.03880328,0.42813387)
          (2.14176758,0.45302272) (2.24406819,0.47817296) (2.34566537,0.50358872) (2.44651787,0.52927418)
          (2.54658288,0.55523365) (2.64581601,0.58147151) (2.74417117,0.60799222) (2.84160054,0.63480037)
          (2.93805451,0.66190062) (3.03348158,0.68929774) (3.12782828,0.71699661) (3.22103915,0.74500221)
          (3.31305659,0.77331963) (3.40382083,0.80195407) (3.49326980,0.83091085) (3.58133906,0.86019542)
          (3.66796172,0.88981332) (3.75306831,0.91977025) (3.83658667,0.95007202) (3.91844188,0.98072458)
          (3.99855612,1.01173401) (4.07684854,1.04310654) (4.15323517,1.07484854) (4.22762873,1.10696652)
          (4.29993858,1.13946715) (4.37007050,1.17235727) (4.43792656,1.20564386) (4.50340499,1.23933409)
          (4.56640000,1.27343527) (4.62680158,1.30795493) (4.68449538,1.34290074) (4.73936245,1.37828059)
          (4.79127910,1.41410254) (4.84011668,1.45037487) (4.88574134,1.48710606) (4.92801381,1.52430478)
          (4.96678918,1.56197996) (5.00191663,1.60014071) (5.03323917,1.63879642) (5.06059338,1.67795668)
          (5.08380909,1.71763135) (5.10270909,1.75783053) (5.11710882,1.79856461) (5.12681601,1.83984423)
          (5.13163035,1.88168031) (5.13134310,1.92408408) (5.12573671,1.96706705) (5.11458438,2.01064107)
          (5.09764968,2.05481826) (5.07468604,2.09961113) (5.04543629,2.14503250) (5.00963215,2.19109554)
          (4.96699372,2.23781382) (4.91722887,2.28520124) (4.86003269,2.33327214) (4.79508683,2.38204124)
          (4.72205887,2.43152370) (4.64060160,2.48173509) (4.55035230,2.53269145) (4.45093194,2.58440930)
          (4.34194439,2.63690561) (4.22297554,2.69019788) (4.09359235,2.74430412) (3.95334195,2.79924290)
          (3.80175054,2.85503331) (3.63832236,2.91169507) (3.46253848,2.96924846) (3.27385567,3.02771442)
          (3.07170502,3.08711452) (2.85549064,3.14747101) (2.62458815,3.20880685) (2.37834321,3.27114571)
          (2.11606982,3.33451205) (1.83704866,3.39893109) (1.54052518,3.46442888) (1.22570770,3.53103231)
          (0.89176531,3.59876918) (0.53782566,3.66766819) (0.16297264,3.73775902) (-0.23375617,3.80907234)
          (-0.65337208,3.88163985) (-1.10555061,3.80061328) (-1.52973564,3.72133342) (-1.92733699,3.64374524)
          (-2.29968671,3.56779594) (-2.64804397,3.49343490) (-2.97359942,3.42061349) (-3.27747954,3.34928506)
          (-3.56075033,3.27940478) (-3.82442106,3.21092956) (-4.06944762,3.14381800) (-4.29673569,3.07803029)
          (-4.50714367,3.01352811) (-4.70148546,2.95027463) (-4.88053302,2.88823435) (-5.04501879,2.82737312)
          (-5.19563793,2.76765803) (-5.33305048,2.70905737) (-5.45788327,2.65154058) (-5.57073183,2.59507819)
          (-5.67216213,2.53964179) (-5.76271217,2.48520395) (-5.84289354,2.43173822) (-5.91319287,2.37921906)
          (-5.97407316,2.32762179) (-6.02597506,2.27692259) (-6.06931804,2.22709845) (-6.10450156,2.17812712)
          (-6.13190610,2.12998709) (-6.15189414,2.08265757) (-6.16481113,2.03611845) (-6.17098633,1.99035027)
          (-6.17073368,1.94533419) (-6.16435260,1.90105199) (-6.15212865,1.85748602) (-6.13433432,1.81461919)
          (-6.11122962,1.77243494) (-6.08306276,1.73091723) (-6.05007069,1.69005050) (-6.01247966,1.64981969)
          (-5.97050580,1.61021019) (-5.92435554,1.57120782) (-5.87422613,1.53279883) (-5.82030606,1.49496989)
          (-5.76277550,1.45770804) (-5.70180669,1.42100073) (-5.63756430,1.38483574) (-5.57020582,1.34920125)
          (-5.49988188,1.31408573) (-5.42673657,1.27947801) (-5.35090774,1.24536724) (-5.27252731,1.21174284)
          (-5.19172152,1.17859456) (-5.10861121,1.14591242) (-5.02331206,1.11368673) (-4.93593481,1.08190804)
          (-4.84658551,1.05056717) (-4.75536572,1.01965520) (-4.66237268,0.98916344) (-4.56769958,0.95908341)
          (-4.47143565,0.92940690) (-4.37366640,0.90012587) (-4.27447376,0.87123253) (-4.17393623,0.84271926)
          (-4.07212903,0.81457865) (-3.96912428,0.78680350) (-3.86499108,0.75938675) (-3.75979568,0.73232157)
          (-3.65360156,0.70560126) (-3.54646962,0.67921930) (-3.43845821,0.65316936) (-3.32962329,0.62744522)
          (-3.22001850,0.60204084) (-3.10969531,0.57695034) (-2.99870302,0.55216795) (-2.88708896,0.52768807)
          (-2.77489847,0.50350521) (-2.66217507,0.47961402) (-2.54896046,0.45600929) (-2.43529464,0.43268592)
          (-2.32121599,0.40963893) (-2.20676129,0.38686346) (-2.09196583,0.36435476) (-1.97686343,0.34210820)
          (-1.86148655,0.32011923) (-1.74586629,0.29838343) (-1.63003248,0.27689647) (-1.51401373,0.25565412)
          (-1.39783747,0.23465224) (-1.28152997,0.21388680) (-1.16511647,0.19335383) (-1.04862110,0.17304947)
          (-0.93206704,0.15296994) (-0.81547649,0.13311154) (-0.69887071,0.11347065) (-0.58227010,0.09404373)
          (-0.46569418,0.07482732) (-0.34916166,0.05581802) (-0.23269047,0.03701253) (-0.11629777,0.01840758)
          (0.00000000,0.00000000)
        } \closedcycle;
        \addplot[black,thick] coordinates {
          (0.00000000,0.00000000) (0.10965277,0.02060405) (0.21917570,0.04140513) (0.32854900,0.06240607)
          (0.43775214,0.08360975) (0.54676382,0.10501910) (0.65556196,0.12663712) (0.76412362,0.14846685)
          (0.87242504,0.17051142) (0.98044153,0.19277398) (1.08814752,0.21525777) (1.19551644,0.23796610)
          (1.30252078,0.26090232) (1.40913195,0.28406986) (1.51532032,0.30747222) (1.62105516,0.33111297)
          (1.72630458,0.35499576) (1.83103554,0.37912430) (1.93521363,0.40350238) (2.03880328,0.42813387)
          (2.14176758,0.45302272) (2.24406819,0.47817296) (2.34566537,0.50358872) (2.44651787,0.52927418)
          (2.54658288,0.55523365) (2.64581601,0.58147151) (2.74417117,0.60799222) (2.84160054,0.63480037)
          (2.93805451,0.66190062) (3.03348158,0.68929774) (3.12782828,0.71699661) (3.22103915,0.74500221)
          (3.31305659,0.77331963) (3.40382083,0.80195407) (3.49326980,0.83091085) (3.58133906,0.86019542)
          (3.66796172,0.88981332) (3.75306831,0.91977025) (3.83658667,0.95007202) (3.91844188,0.98072458)
          (3.99855612,1.01173401) (4.07684854,1.04310654) (4.15323517,1.07484854) (4.22762873,1.10696652)
          (4.29993858,1.13946715) (4.37007050,1.17235727) (4.43792656,1.20564386) (4.50340499,1.23933409)
          (4.56640000,1.27343527) (4.62680158,1.30795493) (4.68449538,1.34290074) (4.73936245,1.37828059)
          (4.79127910,1.41410254) (4.84011668,1.45037487) (4.88574134,1.48710606) (4.92801381,1.52430478)
          (4.96678918,1.56197996) (5.00191663,1.60014071) (5.03323917,1.63879642) (5.06059338,1.67795668)
          (5.08380909,1.71763135) (5.10270909,1.75783053) (5.11710882,1.79856461) (5.12681601,1.83984423)
          (5.13163035,1.88168031) (5.13134310,1.92408408) (5.12573671,1.96706705) (5.11458438,2.01064107)
          (5.09764968,2.05481826) (5.07468604,2.09961113) (5.04543629,2.14503250) (5.00963215,2.19109554)
          (4.96699372,2.23781382) (4.91722887,2.28520124) (4.86003269,2.33327214) (4.79508683,2.38204124)
          (4.72205887,2.43152370) (4.64060160,2.48173509) (4.55035230,2.53269145) (4.45093194,2.58440930)
          (4.34194439,2.63690561) (4.22297554,2.69019788) (4.09359235,2.74430412) (3.95334195,2.79924290)
          (3.80175054,2.85503331) (3.63832236,2.91169507) (3.46253848,2.96924846) (3.27385567,3.02771442)
          (3.07170502,3.08711452) (2.85549064,3.14747101) (2.62458815,3.20880685) (2.37834321,3.27114571)
          (2.11606982,3.33451205) (1.83704866,3.39893109) (1.54052518,3.46442888) (1.22570770,3.53103231)
          (0.89176531,3.59876918) (0.53782566,3.66766819) (0.16297264,3.73775902) (-0.23375617,3.80907234)
          (-0.65337208,3.88163985)
        };
        \addplot[black,thick] coordinates {
          (0.00000000,0.00000000) (-0.11629777,0.01840758) (-0.23269047,0.03701253) (-0.34916166,0.05581802)
          (-0.46569418,0.07482732) (-0.58227010,0.09404373) (-0.69887071,0.11347065) (-0.81547649,0.13311154)
          (-0.93206704,0.15296994) (-1.04862110,0.17304947) (-1.16511647,0.19335383) (-1.28152997,0.21388680)
          (-1.39783747,0.23465224) (-1.51401373,0.25565412) (-1.63003248,0.27689647) (-1.74586629,0.29838343)
          (-1.86148655,0.32011923) (-1.97686343,0.34210820) (-2.09196583,0.36435476) (-2.20676129,0.38686346)
          (-2.32121599,0.40963893) (-2.43529464,0.43268592) (-2.54896046,0.45600929) (-2.66217507,0.47961402)
          (-2.77489847,0.50350521) (-2.88708896,0.52768807) (-2.99870302,0.55216795) (-3.10969531,0.57695034)
          (-3.22001850,0.60204084) (-3.32962329,0.62744522) (-3.43845821,0.65316936) (-3.54646962,0.67921930)
          (-3.65360156,0.70560126) (-3.75979568,0.73232157) (-3.86499108,0.75938675) (-3.96912428,0.78680350)
          (-4.07212903,0.81457865) (-4.17393623,0.84271926) (-4.27447376,0.87123253) (-4.37366640,0.90012587)
          (-4.47143565,0.92940690) (-4.56769958,0.95908341) (-4.66237268,0.98916344) (-4.75536572,1.01965520)
          (-4.84658551,1.05056717) (-4.93593481,1.08190804) (-5.02331206,1.11368673) (-5.10861121,1.14591242)
          (-5.19172152,1.17859456) (-5.27252731,1.21174284) (-5.35090774,1.24536724) (-5.42673657,1.27947801)
          (-5.49988188,1.31408573) (-5.57020582,1.34920125) (-5.63756430,1.38483574) (-5.70180669,1.42100073)
          (-5.76277550,1.45770804) (-5.82030606,1.49496989) (-5.87422613,1.53279883) (-5.92435554,1.57120782)
          (-5.97050580,1.61021019) (-6.01247966,1.64981969) (-6.05007069,1.69005050) (-6.08306276,1.73091723)
          (-6.11122962,1.77243494) (-6.13433432,1.81461919) (-6.15212865,1.85748602) (-6.16435260,1.90105199)
          (-6.17073368,1.94533419) (-6.17098633,1.99035027) (-6.16481113,2.03611845) (-6.15189414,2.08265757)
          (-6.13190610,2.12998709) (-6.10450156,2.17812712) (-6.06931804,2.22709845) (-6.02597506,2.27692259)
          (-5.97407316,2.32762179) (-5.91319287,2.37921906) (-5.84289354,2.43173822) (-5.76271217,2.48520395)
          (-5.67216213,2.53964179) (-5.57073183,2.59507819) (-5.45788327,2.65154058) (-5.33305048,2.70905737)
          (-5.19563793,2.76765803) (-5.04501879,2.82737312) (-4.88053302,2.88823435) (-4.70148546,2.95027463)
          (-4.50714367,3.01352811) (-4.29673569,3.07803029) (-4.06944762,3.14381800) (-3.82442106,3.21092956)
          (-3.56075033,3.27940478) (-3.27747954,3.34928506) (-2.97359942,3.42061349) (-2.64804397,3.49343490)
          (-2.29968671,3.56779594) (-1.92733699,3.64374524) (-1.52973564,3.72133342) (-1.10555061,3.80061328)
          (-0.65337208,3.88163985)
        };
        \addplot[black,only marks,mark=*,mark size=2.1pt]
          coordinates {(0,0) (-0.65337208,3.88163985)};
        \addplot[black,only marks,mark=square*,mark size=2.5pt]
          coordinates {(5.13214525,1.90054405)};
        \node[surplus label,anchor=north west]
          at (axis cs:0.18,0.3) {No disclosure};
        \node[surplus label,anchor=south west]
          at (axis cs:-0.55,3.91) {Full disclosure};
        \node[surplus label,anchor=south east,align=right]
          at (axis cs:5.18,1.97) {Consumer-optimal\\policy};
      \end{axis}
    \end{tikzpicture}
    \par\smallskip
    {\small (b) $\Omega=\{0.8,1\}$}
  \end{minipage}
  \caption{Attainable surplus changes in Example~\ref{ex:beta33}. For any information policy, $\Delta\CS$ is expected consumer surplus minus its no-disclosure value, and $\Delta\PS$ is expected producer surplus minus its no-disclosure value. The shaded set is the full attainable region over all information policies. In panel (a), both axes are multiplied by $10^3$. In panel (b), the horizontal axis is multiplied by $10^5$ and the vertical axis by $10^3$.}
  \label{fig:beta33}
\end{figure}
\FloatBarrier

\section{Conclusion}

\textcite{schlee1996value} showed that information can hurt consumers when a monopolist adjusts its price in response. We provide a necessary and sufficient condition for information to benefit consumers instead, expressed in terms of Myerson's virtual type. The condition requires a weighted combination of its slope and curvature to be nonnegative.
Producer surplus always rises, while a weaker condition governs total surplus.

The positive result extends well beyond linear demand.
Information benefits consumers whenever the virtual type is convex. This condition is satisfied by a broad class of nonlinear demand curves, including those induced by Beta$(\alpha,\beta)$ type distributions with $\alpha\leq 1$.
Even when convexity fails, the same conclusion holds if monopoly output is sufficiently small.
The Beta$(3,3)$ example also shows the boundary: full disclosure can reduce consumer surplus. The policy lesson is not that information should be hidden, but that disclosure and the monopolist's price response should be evaluated together.

\clearpage

\begin{appendices}

\section{Proofs}

\subsection{Preliminaries}

\begin{lemma}[The nonnegative range of the virtual type]\label{lem:tail}
There is a unique $\theta_0\in(0,\bar\theta)$ such that $\varphi(\theta_0)=0$. The virtual type is negative below $\theta_0$, strictly increasing above $\theta_0$, and
\[
  \lim_{\theta\uparrow\bar\theta}r(\theta)=0,
  \qquad
  \lim_{\theta\uparrow\bar\theta}\varphi(\theta)=\bar\theta.
\]
\end{lemma}

\begin{proof}
Write $S=1-F$. Since $S'=-f$ and $r=S/f$,
\[
  S(y)=S(x)\exp\!\left(-\int_x^y\frac{\dd u}{r(u)}\right),
  \qquad 0<x<y<\bar\theta.
\]
First, $\varphi$ is negative near zero. Otherwise, for every $y>0$ there would be $x<y$ with $\varphi(x)\ge0$. Regularity implies that $\varphi$ is then positive and strictly increasing on $[x,y]$, so $r(u)\le u$ for all $u>0$. The display would give $S(y)\le S(x)x/y$; letting $x\downarrow0$ yields $S(y)=0$, a contradiction.

Because $S(\theta)\to0$ as $\theta\uparrow\bar\theta$, the display also implies $\liminf_{\theta\uparrow\bar\theta}r(\theta)=0$. Hence $\varphi(\theta)>0$ along a sequence approaching $\bar\theta$. Continuity gives a zero, and regularity makes it unique and makes $\varphi$ strictly increasing above it.

On this nonnegative range, $r'=1-\varphi'<1$. If $r$ did not converge to zero, some $\varepsilon>0$ and $x_n\uparrow\bar\theta$ would satisfy $r(x_n)\ge\varepsilon$. For $y$ close enough to $\bar\theta$, choose $x_n>y$. Then $x_n-y<\varepsilon/2$, and integration of $r'<1$ gives
\[
  r(y)>r(x_n)-(x_n-y)>\frac{\varepsilon}{2},
\]
contradicting $\liminf r=0$. Thus $r\to0$, and $\varphi(\theta)=\theta-r(\theta)\to\bar\theta$.
\end{proof}

\begin{lemma}[Posterior demand pairs]\label{lem:pairs}
Every posterior demand pair lies in $K$. Conversely, every point of $K$ is a limit of posterior pairs generated by positive-probability messages under $\mu_0$.
\end{lemma}

\begin{proof}
Only the converse needs proof. Write $A(\omega)=(a(\omega),b(\omega))$ and fix $k\in K$. By Carath\'eodory's theorem,
\[
  k=\sum_{j=1}^m\lambda_jA(\omega_j),
  \qquad m\le3,
  \qquad \lambda_j\ge0,
  \qquad \sum_{j=1}^m\lambda_j=1.
\]
For $\delta>0$, choose neighborhoods $U_j$ of $\omega_j$ with positive prior probability such that $\lVert A(\omega)-A(\omega_j)\rVert<\delta$ on $U_j$. Then
\[
  \nu_\delta=\sum_{j=1}^m\lambda_j\mu_0(\,\cdot\mid U_j)
\]
has a bounded density $h_\delta$ relative to $\mu_0$, and its demand pair is within $\delta$ of $k$. Choose $\varepsilon>0$ with $\varepsilon h_\delta\le1$. Sending a message with probability $\varepsilon h_\delta(\omega)$ in state $\omega$ produces posterior $\nu_\delta$ after a positive-probability message. Let $\delta\downarrow0$.
\end{proof}

\begin{lemma}[Information and convexity]\label{lem:information-convexity}
Let $H:K\to\mathbb R$ be continuous. Its expected value is increasing in information if and only if $H$ is convex on $K$.
\end{lemma}

\begin{proof}
If $H$ is convex, the posterior pair after a garbling is the conditional mean of the pair before the garbling, so conditional Jensen's inequality proves sufficiency.

For necessity, suppose that, for some $k_1,k_2\in K$ and $\lambda\in(0,1)$,
\[
  H\bigl(\lambda k_1+(1-\lambda)k_2\bigr)
  >\lambda H(k_1)+(1-\lambda)H(k_2).
\]
By Lemma~\ref{lem:pairs} and continuity, the same strict inequality holds for nearby pairs induced by beliefs $\nu_i=h_i\mu_0$ with bounded densities. Choose $\varepsilon>0$ such that
\[
  \varepsilon\bigl[\lambda h_1+(1-\lambda)h_2\bigr]\le1 \qquad
\mu_0\text{-a.e.}
\]
Send message $1$ with probability $\varepsilon\lambda h_1(\omega)$, message $2$ with probability $\varepsilon(1-\lambda)h_2(\omega)$, and a residual message otherwise. Pooling messages $1$ and $2$ induces $\lambda k_1+(1-\lambda)k_2$. Revealing which one occurred changes expected $H$ by
\[
  \varepsilon\!\left[
    \lambda H(k_1)+(1-\lambda)H(k_2)
    -H\bigl(\lambda k_1+(1-\lambda)k_2\bigr)
  \right]<0.
\]
Thus monotonicity in information fails.
\end{proof}

\begin{lemma}[One-dimensional reduction]\label{lem:reduction}
Let $I=\{t(a,b):(a,b)\in K\}$ and let $g:I\to\mathbb R$ be continuous. The function
\[
  H(a,b)=a\,g\!\left(t(a,b)\right)
\]
is convex on $K$ if and only if $g$ is convex on $I$.
\end{lemma}

\begin{proof}
For $(a_i,b_i)\in K$ and $\lambda\in(0,1)$, put
\[
  (\bar a,\bar b)=\lambda(a_1,b_1)+(1-\lambda)(a_2,b_2),
  \qquad
  \eta=\frac{\lambda a_1}{\bar a}.
\]
Then
\[
  t(\bar a,\bar b)=\eta t(a_1,b_1)+(1-\eta)t(a_2,b_2).
\]
This identity proves that convexity of $g$ implies convexity of $H$. Conversely, for pairs inducing arbitrary $t_1,t_2\in I$ and any $\eta\in(0,1)$, choose
\[
  \lambda=\frac{\eta a_2}{(1-\eta)a_1+\eta a_2}.
\]
Dividing the convexity inequality for $H$ by $\bar a$ gives the convexity inequality for $g$ with weight $\eta$.
\end{proof}

\begin{lemma}[Surplus curvature]\label{lem:curvature}
Consumer surplus is convex on $K$ if and only if condition~\eqref{eq:cs-condition} holds. Total surplus is convex on $K$ if and only if condition~\eqref{eq:ts-condition} holds.
\end{lemma}

\begin{proof}
Let $I=\{t(a,b):(a,b)\in K\}$ and let $x(t)$ be the cutoff in \eqref{eq:cutoff}. Define
\[
  g_C(t)=\int_{x(t)}^{\bar\theta}(\theta-x(t))f(\theta)\dd\theta,
  \qquad
  g_T(t)=\int_{x(t)}^{\bar\theta}(\theta-t)f(\theta)\dd\theta.
\]
Then $\CS(a,b)=a g_C(t(a,b))$ and $\TS(a,b)=a g_T(t(a,b))$, so Lemma~\ref{lem:reduction} reduces the claim to convexity of $g_C$ and $g_T$ on $I$.

If $x_L=x_H$, the conditions are vacuous. The two functions are either identically zero or evaluated at a single $t$, and are therefore convex. Suppose $x_L<x_H$. At a trading value $t<\bar\theta$, write $x=x(t)$ and $p=\varphi'(x)>0$. Since $t=\varphi(x)$,
\[
  x'(t)=\frac1p,
  \qquad
  1-F(x)=f(x)r(x).
\]
Leibniz's rule gives
\[
  g_C'(t)=-\frac{1-F(x)}p,
  \qquad
  g_T'(t)=-(1-F(x))\left(1+\frac1p\right),
\]
and hence
\[
  g_C''(t)
  =\frac{f(x)}{p^3}\bigl[p+r(x)\varphi''(x)\bigr],
\]
\[
  g_T''(t)
  =\frac{f(x)}{p^3}\bigl[p+p^2+r(x)\varphi''(x)\bigr].
\]
The prefactors are positive. Since $K$ is connected and $x(a,b)$ is continuous, the trading cutoffs fill $(x_L,x_H)$. Thus the two displayed curvature conditions are necessary and sufficient on the trading part of $I$.

If $I$ also contains $t\ge\bar\theta$, then $g_C=g_T=0$ there. On the trading side both functions decrease to zero as $t\uparrow\bar\theta$. Under the displayed conditions their left slopes are nondecreasing and at most zero, while the constant branch has slope zero. The extension is therefore convex.
\end{proof}

\subsection{Main Results}

\begin{proof}[Proof of Lemma~\ref{lem:cutoff}]
Fix $(a,b)\in K$ and write $t=t(a,b)$. Apart from the positive factor $a$, the objective in \eqref{eq:ps} is $(x-t)[1-F(x)]$, whose interior derivative is
\[
  f(x)[t-\varphi(x)].
\]
If $t<\bar\theta$, Lemma~\ref{lem:tail} shows that the derivative changes sign once, at $x=\varphi^{-1}(t)$, which is therefore the unique maximizer. If $t\ge\bar\theta$, every $x<\bar\theta$ has a negative margin, whereas $x=\bar\theta$ gives zero profit, so no trade is uniquely optimal. Continuity follows from continuity of $t$ and $\varphi^{-1}$, including $\varphi^{-1}(t)\uparrow\bar\theta$ as $t\uparrow\bar\theta$.
\end{proof}

\begin{proof}[Proof of Proposition~\ref{prop:producer}]
For each cutoff $x$, $(ax+b-c)[1-F(x)]$ is affine in $(a,b)$. Its maximum over the compact cutoff set is therefore convex and continuous. Lemma~\ref{lem:information-convexity} proves the result.
\end{proof}

\begin{proof}[Proof of Theorem~\ref{thm:consumer}]
Consumer surplus is continuous on $K$. Lemmas~\ref{lem:curvature} and~\ref{lem:information-convexity} give the stated equivalence.
\end{proof}

\begin{proof}[Proof of Theorem~\ref{thm:total}]
Total surplus is continuous on $K$. Lemmas~\ref{lem:curvature} and~\ref{lem:information-convexity} give the stated equivalence.
\end{proof}

\begin{proof}[Proof of Corollary~\ref{cor:convex-virtual}]
At every $\theta\in(x_L,x_H)$, regularity gives $\varphi'(\theta)>0$, while convexity gives $\varphi''(\theta)\ge0$. Conditions~\eqref{eq:cs-condition} and~\eqref{eq:ts-condition} follow. Apply Proposition~\ref{prop:producer} and Theorems~\ref{thm:consumer}--\ref{thm:total}.
\end{proof}

\begin{proof}[Proof of Proposition~\ref{prop:small-output}]
Choose $\ell_0<\bar\theta$ and $M\ge0$ such that $\varphi''\ge-M$ on $(\ell_0,\bar\theta)$. Since $r''=-\varphi''\le M$, the function $r'(\theta)-M\theta$ is nonincreasing; hence $r'$ has an extended limit at $\bar\theta$. Lemma~\ref{lem:tail} gives $r(\theta)\to0$, and positivity of $r$ rules out a positive limit of $r'$. Therefore, for some $\ell\in(\ell_0,\bar\theta)$,
\[
  r'(\theta)\le\frac12,
  \qquad
  Mr(\theta)\le\frac12
  \qquad
  (\ell\le\theta<\bar\theta).
\]
Using $\varphi'=1-r'$ and $\varphi''\ge-M$,
\[
  \varphi'(\theta)+r(\theta)\varphi''(\theta)
  \ge1-r'(\theta)-Mr(\theta)\ge0.
\]
Condition~\eqref{eq:ts-condition} follows because it adds $(\varphi')^2$. Set $\bar q=1-F(\ell)>0$. If every posterior output is at most $\bar q$, strict decrease of $1-F$ implies that every optimal cutoff is at least $\ell$. Theorems~\ref{thm:consumer} and~\ref{thm:total} complete the proof.
\end{proof}

\subsection{Verification of the Examples}

\begin{proof}[Verification of Example~\ref{ex:beta-family}]
Let $\gamma=1-\alpha\in[0,1)$. A change of variables in $r=(1-F)/f$ gives
\[
  r(\theta)
  =(1-\theta)\int_0^1 y^{\beta-1}
  \left(\frac{\theta}{1-(1-\theta)y}\right)^\gamma\dd y.
\]
For fixed $y$, define
\[
  j_y(\theta)=(1-\theta)
  \left(\frac{\theta}{1-(1-\theta)y}\right)^\gamma.
\]
Direct differentiation yields
\[
  j_y''(\theta)
  =-\frac{\gamma(1-y)\theta^{\gamma-2}}
  {(1-y+y\theta)^{\gamma+2}}
  \left[(1+\theta)-(1-\theta)\{\gamma(1-y)+y\}\right]\le0,
\]
because $\gamma(1-y)+y\le1$. Hence $r$ is concave and $\varphi(\theta)=\theta-r(\theta)$ is convex.

If $\alpha<1$, then $r(\theta)\sim \mathrm B(\alpha,\beta)\theta^{1-\alpha}$ as $\theta\downarrow0$, so $r(\theta)/\theta\to\infty$ and $\varphi$ is negative near zero. The preceding inequality is strict after integration, so $\varphi$ is strictly convex; moreover the integral formula gives $r(\theta)\to0$ as $\theta\uparrow1$. Thus $\varphi$ crosses zero once, with positive derivative there, and $\varphi'>0$ wherever $\varphi\ge0$. If $\alpha=1$, then
\[
  \varphi(\theta)=\frac{(\beta+1)\theta-1}{\beta},
\]
which is strictly increasing. The distribution is therefore regular, and Corollary~\ref{cor:convex-virtual} applies.
\end{proof}

\begin{proof}[Verification of Example~\ref{ex:beta33}]
Write $z=\E[\omega\mid s]$. The posterior demand pair is $(1,z)$ and the cutoff $x(z)$ solves $\varphi(x)=1-z$. For Beta$(3,3)$,
\[
  f(x)=30x^2(1-x)^2,
  \qquad
  S(x)=1-F(x)=(1-x)^3(6x^2+3x+1),
\]
\[
  r(x)=\frac{(1-x)(6x^2+3x+1)}{30x^2},
  \qquad
  \varphi(x)=\frac{36x^3-3x^2-2x-1}{30x^2}.
\]
Moreover,
\[
  \varphi'(x)=\frac{(3x+1)(6x^2-2x+1)}{15x^3}>0,
  \qquad
  \varphi''(x)=-\frac{2x+3}{15x^4},
\]
so the distribution is regular and $\varphi''$ is bounded below near one. The left side of \eqref{eq:cs-condition} is
\[
  \frac{540x^6+42x^4+42x^3-13x^2-8x-3}{450x^6}.
\]
Its numerator has exactly one positive root: it is negative at zero, positive at one, and has one sign change. Numerically,
\[
  x_c\approx0.4395375,
  \qquad
  S(x_c)\approx0.6122669,
\]
and condition~\eqref{eq:cs-condition} holds exactly for $x\ge x_c$. With states $\{0.6,0.8\}$, the smallest cutoff is
\[
  x(0.8)\approx0.4836313,
  \qquad
  S(x(0.8))\approx0.5306694<S(x_c).
\]
Thus Proposition~\ref{prop:small-output} applies with $\bar q=S(x_c)$.

For completeness, define posterior consumer and producer surplus by
\[
  C(z)=\CS(1,z),
  \qquad
  P(z)=\PS(1,z).
\]
Integration and the first-order condition give
\[
  C(z)=\frac{(1-x(z))^4(2x(z)^2+2x(z)+1)}2,
  \qquad
  P(z)=r(x(z))S(x(z)).
\]
For either binary state space, an information policy induces a distribution $\tau$ of $z$ supported on $[\omega_L,\omega_H]$ with mean $(\omega_L+\omega_H)/2$. Conversely, every such $\tau$ is feasible. Indeed, with
\[
  p(z)=\frac{z-\omega_L}{\omega_H-\omega_L},
\]
define the conditional signal measures by $2p(z)\tau(\dd z)$ in the high state and $2[1-p(z)]\tau(\dd z)$ in the low state. Bayes' rule then gives posterior mean $z$.

Now take states $\{0.8,1\}$. By Lemma~\ref{lem:curvature}, $C$ is convex below
\[
  z_c=1-\varphi(x_c)\approx0.8967684
\]
and concave above it. Let
\[
  D(z)=(z-0.8)C'(z)-C(z)+C(0.8).
\]
Then $D'(z)=(z-0.8)C''(z)$, so $D$ first increases and then decreases. Direct evaluation gives
\[
  D(0.9)\approx2.19\times10^{-4}>0,
  \qquad
  D(1)\approx-8.63\times10^{-4}<0.
\]
Hence there is a unique $z^*\in(0.9,1)$ such that
\[
  C'(z^*)=\frac{C(z^*)-C(0.8)}{z^*-0.8},
  \qquad
  z^*\approx0.9475433.
\]
Because $D(z)>0$ for $0.8<z<z^*$, the chord from $0.8$ to $z^*$ lies above $C$; joined to $C$ on $[z^*,1]$, it is the concave envelope of $C$. At the prior mean $0.9$, the maximizing distribution therefore puts probability
\[
  \lambda^*=\frac{z^*-0.9}{z^*-0.8}\approx0.3222327
\]
on posterior mean $0.8$ and the remaining probability on $z^*$. Since the low state has prior probability one half, this is implemented by sending $\mathsf B$ in the low state with probability
\[
  \rho^*=2\lambda^*\approx0.6444653,
\]
and always sending $\mathsf G$ in the high state. Bayes' rule gives posterior mean $z^*$ after $\mathsf G$.

Substitution gives, relative to no disclosure,
\[
\begin{array}{c|cc}
 & \Delta\CS & \Delta\PS\\ \hline
\text{full disclosure}
 & -6.53\times10^{-6} & 3.88\times10^{-3}\\
\text{consumer-optimal policy}
 & 5.13\times10^{-5} & 1.90\times10^{-3}.
\end{array}
\]
Finally, the extreme distributions on an interval with fixed mean have at most two support points. Expected surplus is affine in $\tau$, so the closed convex hull of the surplus pairs generated by these two-point distributions is the full attainable region. Evaluating this characterization gives both panels of Figure~\ref{fig:beta33}.
\end{proof}

\end{appendices}

\begingroup
\singlespacing
\footnotesize
\setlength{\bibitemsep}{0pt}
\printbibliography[title={References}]
\endgroup

\end{document}